\documentclass{aa}  

\usepackage{graphicx}
\usepackage{nicefrac}
\usepackage{threeparttable}
\usepackage{txfonts}
\usepackage{float}
\usepackage{comment}
\usepackage{geometry}
\usepackage{arydshln}
\usepackage{xcolor}
\usepackage{hyperref}
\begin{document}

   \title{Impact of Population~III homogeneous stellar evolution on early cosmic reionisation}

   \author{Y. Sibony
          \inst{1}
        \and
          B. Liu
          \inst{2}
          \and
          C. Simmonds
          \inst{1}
          \and
          G. Meynet
          \inst{1}
          \and
          V. Bromm
          \inst{2}}

   \institute{Observatoire de Genève,
              Chemin Pegasi 51, 1290 Versoix, Switzerland\\
              \email{yves.sibony@unige.ch}
         \and
             Department of Astronomy, University of Texas, Austin, TX 78712, USA\\
             }

   \date{}

 
  \abstract
   {Population~III (Pop~III) stars may be fast rotating. An expected  consequence of fast rotation is strong internal mixing that deeply affects their evolutionary tracks in the Hertzsprung-Russell diagram and hence their ionising power.}
   {We investigate the impact on the ionising power of Pop III stars in an extreme case of internal mixing, the one leading to chemically homogeneous evolution (CHE). In that situation, during the main sequence phase, the star keeps the same chemical composition from its centre to its surface. Homogeneous stars have larger effective temperatures and luminosities than stars evolving non-homogeneously and thus are stronger ionising sources.
   }
   {The stellar evolution models are based on $n=3$ polytropes with a time-varying mass fraction of hydrogen. The ionisation model employs the self-similar champagne flow solution from \citet{Shu2002} and numerical simulations for the stochastic treatment of star clusters over a grid of redshifts and halo masses.}
   {We find that haloes containing chemically homogeneous stars have an escape fraction of ionising photons up to twice that of haloes containing classical Pop~III stars. 
   By extrapolating the high-$z$ ionisation history powered by Pop~III stars (at $z\gtrsim 15$) to the post-reionisation epoch, we derive the Thomson scattering optical depth $\tau$, which is compared with the value measured by \textit{Planck}. We find that $\tau$ is overproduced by $\sim1.5- 5\sigma$ when all Pop III stars evolve homogeneously. This indicates that CHE is unlikely to be realised in the majority of Pop~III stars, although the present study cannot exclude that a fraction of them undergo CHE.}
   {Fast rotation might have a significant impact on the ionising budget of Pop III stars and thus on early cosmic reionisation. The impact is stronger for less top-heavy initial mass functions of Pop~III stars.}


   \maketitle
%

\section{Introduction}

The stars we know from our local Universe formed from giant molecular clouds, which have experienced chemical enrichment (or pollution, depending on one's point of view) from previous stellar generations. As a result all the stars we have ever observed contain at least traces of metals. Since all the known mechanisms that produce elements heavier than beryllium are stellar (or of stellar origin, e.g. supernovae or neutron star mergers), there must have been a series of generations of stars, each one forming from the ejecta of the previous one, being more and more metal-rich as time went on. Logically then, stars without any metals---the so-called Population~III (Pop III) stars--- while not having been observed yet, must have existed as the first generation of stars, forming from primordial Big Bang nucleosynthesis material around a redshift of $z\sim30-10$ \citep[e.g.][]{Barkana2001}.

Because the Universe was very different then than it is now (chemically, but also in terms of average density and temperature, as well as the intensity of magnetic fields), the processes of Pop~III star formation \citep[e.g.][]{Abel2000, Bromm2013} and stellar evolution \citep[see e.g.][]{Heger2010, Yoon2012, Murphy2021} were also very different from what we can observe in the local Universe \citep[for a review article concerning Pop III stars, see][]{Bromm2017}. A number of studies on the theoretical properties of these stars have been conducted in recent decades, concluding---among many other things---that Pop~III stars should be more massive and their initial mass function (IMF) more top-heavy than current stars \citep{Greif2011,Stacy2013,Susa2014,Wollenberg2020}. Furthermore, for the same mass, a Pop~III star should exhibit a larger effective temperature and luminosity than its metal-enriched counterpart \citep{Ekstrom2008,Yoon2012,Murphy2021}. 
Also, Pop III stars may be fast rotators because they rapidly accrete gas with high angular momentum, which is typical in hot, thick Pop~III star-forming disks with inefficient cooling \citep{Stacy2013}. Rotational mixing renders stars more luminous and hotter during their main sequence phase, thus enhancing their capacity to produce ionising photons.
This implies that their production rate of ionising photons could be very large; indeed Pop~III stars are expected to be one of the causes \citep[along with active galactic nuclei and Population II stars around $z\sim20-15$, although the latter is expected to exhibit much softer spectra than their metal-free counterparts, see][]{Schaerer2003} of the early reionisation period ($z\sim 30-15$) which followed the dark ages of the Universe.

Very fast rotation may induce such strong mixing inside stars so that they follow a homogeneous chemical evolution, representing an extreme limit of what single star evolution can achieve in terms of ionising power. 
In order to place upper limits on the ionising flux emitted by the first stars, one can consider the case where all Pop III stars would follow a chemically homogeneous evolution (CHE) path. Indeed, stars evolving under this constraint of being perfectly mixed at all times would be even more compact and would thus have an even hotter surface. Furthermore, their bluewards (instead of the classical redwards) evolution  during the MS, coupled with their larger fuel supply compared to standard models, would make chemically homogeneous stars the largest emitters of integrated ionising radiation over the population's lifetime. We propose an analytical model including chemically homogeneous stellar evolution and ionising radiation in order to quantify the aforementioned upper limit. We then derive upper limits on the reionisation of the intergalactic medium (IGM) by Pop~III stars. We compare our results to those obtained with classical non-rotating Pop~III models computed with the Geneva stellar evolution code (GENEC).

This article is organised as follows. In section \ref{sec:CHE} we detail the analytical model for homogeneous evolution. In section \ref{sec:ioni} we present the ionisation model, based on the self-similar champagne flow solution \citep{Shu2002}. Section \ref{sec:results} presents our results and compares them to those obtained with the standard Pop III models. 
Section \ref{sec:discussion} discusses some limitations of the present approach. Finally, we summarise our main findings in section \ref{sec:conclusions}.

\section{Chemically homogeneous stellar models}
\label{sec:CHE}
When the matter processed by nucleosynthesis in the stellar core is mixed with the whole stellar mass on timescales much shorter than the nuclear timescale, stars have a chemically homogeneous evolution. This type of evolution has been advocated for explaining young blue stragglers in young open star clusters \citep{Maeder1987}, the evolution of long soft Gamma Ray Bursts \citep{Yoon2005}, and the evolution leading to the formation of close binary black holes \citep{Mandel2016, deMink2016, Marchant2017, Buisson2020}. At the moment, the observational evidence for stars following such an evolution remains scarce. However, as discussed in \citet{Martins2009, Marchant2017}, some early-type massive stars appear to be very good candidates. The physical mechanism responsible for the strong mixing is probably linked to the hydro- or magnetohydrodynamical instabilities triggered by a fast rotation. For now however, the physics of rotation still remain a debated question and different hypotheses lead to different results \citep[see e.g. the discussion in][]{Song2016}. In the present work, we do not focus on the physical cause of homogeneous evolution but rather on the consequences such an evolution would have on cosmic reionisation. This is why we use here a very simple approach with an analytical model for such stars.

The quantities we aim to derive from our analytical model are the star's luminosity and effective temperature. Such models have been made in the literature with polytropes \citep{BeechMitalas1989}. While this could also be a physically relevant approach for advanced phases of evolution, we focus here on the MS phase. We do not consider mass loss as the stars do not reach the Eddington luminosity, and at primordial chemical composition radiatively line-driven winds should be negligible.

A polytrope model is characterised by a global relation between pressure and density:
\begin{equation}
    P(r) = K \rho(r)^{\frac{n+1}{n}}.
\end{equation}
Here, we consider the special case $n=3$ (the so-called Eddington Standard model). In this case, the Eddington quartic relation provides a relation between the mass $m_\star$ of the star, its mean molecular weight $\mu$ (constant throughout the star by virtue of chemical homogeneity), and its ratio of gas to total pressure, $\beta$:
\begin{equation}
\label{eqn2}
    \frac{1-\beta}{\beta^4} = A~ \mu^4 \left(\frac{m_\star}{M_\odot}\right)^2,
\end{equation}
where $A = 0.002986$ is a constant. Equation~(\ref{eqn2}) has four solutions for $\beta$, two of which are complex, one is a negative real number, and one is a positive real number. Hence for a given stellar mass (we do not consider the case of mass loss), at any stage of its MS phase (since $\mu$ is directly linked to the hydrogen mass fraction $X$ by $\mu = (0.75+1.25X)^{-1}$ for a Pop III star), we can compute the value of $\beta$.\\
Given the three quantities $m_\star$, $X$, and $\beta$, we use Eq.~(5.47) of \citet{Prialnik2009} to compute the luminosity:
\begin{equation}
\label{eqn3}
    \frac{L}{L_\odot} = 0.003~ \frac{4 \pi c G M_\odot}{\kappa_s L_\odot}~ \mu^4 \beta^4 \left(\frac{m_\star}{M_\odot}\right)^3,
\end{equation}
where $\kappa_s$ is the surface electron scattering opacity assuming that the medium is completely ionised:
\begin{equation}
    \kappa_s = 0.2~(1+X)~\rm{cm^2/g}.
\end{equation}
We note that the mass-luminosity relation simplifies to $L\propto m_\ast$ for very massive stars ($m_\ast \gtrsim 100~M_\odot$; \citealt{BKL2001}).
Our next quantity of interest is the star's effective temperature. Here we use the following relation between the central and effective temperature of a homogeneous $n=3$ polytrope from \citet{HoyleFowler1963}:
\begin{equation}
\label{eqn5}
    \left(\frac{T_{\rm eff}}{10^4 \rm{K}}\right) = 1.010~ (1-\beta)^{1/8} (1+X)^{-1/4} \left(\frac{T_c}{10^6 \rm{K}}\right),
\end{equation}
where $T_c$ is the star's central temperature.\\
The system of Eqs.~(\ref{eqn3}) and (\ref{eqn5}) is not fully constrained by the sole choice of a stellar mass, which determines only the luminosity on the zero-age main sequence (ZAMS). We also need to fix an initial value for the radius, or equivalently, effective temperature, central temperature, central density, or mean density. To that end, we extract the central temperature on the ZAMS from the Pop III models of \citet{Murphy2021}. This initial condition constrains the initial effective temperature and thus the star's initial radius. We then evolve the central temperature using the following relation (which is just a consequence of the perfect gas equation of state and of the hydrostatic equilibrium equation) :
\begin{equation}
\label{eqn6}
   T_c \propto \frac{\beta \mu m_\star }{R_\star},
\end{equation}
though it turns out that the change in $T_c$ over the MS for the CHE models ends up being negligible.\\
The method goes as follows: for the masses at which \citet{Murphy2021} computed Pop III models (complemented by models at 180, 250, and 300 $M_\odot$ computed with the same method), we vary $X$ from its initial primordial value of $X=0.75$ down to $X=0$ (as hydrogen is burned by the star), and compute the evolution of $\beta$, $\mu$, and $\kappa_s$. This allows us to compute the evolution of the luminosity with Eq.~(\ref{eqn3}) and of the central temperature with Eq.~(\ref{eqn6}). Then, using Eq.~(\ref{eqn5}), we can get to the effective temperature. As a result, the stars' evolution in the HR diagram can be plotted, and their radius computed from $L$ and $T_{\rm eff}$.

In our model, the mass fraction of hydrogen $X$ serves as a proxy for time. However, in the halo ionisation model that we shall introduce further on, we need the actual time evolution of the ionising radiation. In order to go from $X$ to $t$, we use the equivalence of mass and energy and the fact that hydrogen burning generates 0.007 grams of energy for each gram of hydrogen consumed. Thus in the time $dt$, the mass $m_\star~ dX$ of hydrogen is consumed, generating a luminosity
\begin{equation}
    L(t) = 0.007~\frac{dX}{dt} m_\star c^2.
\end{equation}
Inverting the above relation provides us with the duration of each timestep, given the variation of hydrogen and the luminosity at that timestep. This allows us to express the surface quantities of stars as functions of time rather than hydrogen fraction. We also obtain the lifetime of each star by equating it to the time at which the star reaches $X=0$.

Finally, we consider the stars to be perfect black bodies of temperature $T_{\rm eff}$. The rate of photon production as a function of time and stellar mass is given by the integral over the relevant range of wavelengths of the black-body spectral energy distribution, multiplied by the surface of the star:
\begin{equation}
\label{eqn8}
    \dot{N}_{\gamma}(\Delta\lambda,t,m_\star) = \int_{\Delta\lambda} d\lambda F_\lambda(T_{\rm eff}(m_\star, t))\frac{\lambda}{hc} ~ 4 \pi ~R^{2}_{\star}(m_\star, t),
\end{equation}
where $F_\lambda(T_{\rm eff}(m_\star, t))=\pi B_\lambda(T_{\rm eff}(m_\star, t))$ is the outgoing spectral flux density for the black body, $\Delta\lambda$ is the wavelength range of the photons we consider, and $R_\star(m_\star, t)$ is the stellar radius for the star of mass $m_\star$ at time $t$. 
\section{Ionisation model}
\label{sec:ioni}

Once $\dot{N}_{\gamma}(\Delta\lambda,t,m_{\star})$ is known, one can calculate the reionisation history (at $z\gtrsim 15$, dominated by Pop III stars) in two steps. The first step is to calculate the (average) escaping ionising photon number $Q_{\rm esc}(\Delta\lambda,M_h,z)$ per Pop~III star-forming halo 
for different halo masses $M_h$ and redshifts $z$, in the wavelength ranges of interest. The second step is to calculate the evolution of the IGM ionisation fraction $Q_{\rm ion}(z)$ by combining the contribution of all haloes hosting Pop III stars.

\subsection{Ionising photons}
\label{subsubsec:ionimodel}
We consider three types of ionising photons: those capable of ionising H\textsc{i} ($E > 13.6$ eV, thereafter called `H\textsc{ii} photons'), those capable of ionising He\textsc{i} ($E > 24.6$ eV), thereafter `He\textsc{ii} photons', and those capable of ionising He\textsc{ii} ($E > 54.4$ eV, thereafter `He\textsc{iii} photons'). Since computing the interactions of the produced photons with the surrounding medium properly with radiative transfer is beyond the scope of this analytical work, we instead consider an idealised scenario where photons ionise the species with the highest ionisation potential possible:
\begin{align*}
\begin{cases}
    13.6\ \rm{eV} < & E_\gamma < 24.6\ \rm{eV: H\textsc{i} \xrightarrow{} H\textsc{ii},}\\
    24.6\ \rm{eV} < & E_\gamma < 54.4\ \rm{eV: He\textsc{i} \xrightarrow{} He\textsc{ii},}\\
    &E_\gamma > 54.4\ \rm{eV: He\textsc{ii} \xrightarrow{} He\textsc{iii}.}
\end{cases}
\end{align*}
We also compute the number of escaping Lyman-Werner (LW, $11.2\ {\rm eV} < E_\gamma < 13.6\ \rm{eV}$) photons. 
These photons can dissociate H$_2$ molecules and thus are important to produce atomic-cooling haloes in which one expects the possible formation of very and even supermassive stars \citep[see e.g. the review by][and references therein]{Lionel2020}. 
We consider that those photons only escape if hydrogen is fully ionised \citep[see e.g.][]{Schauer2015,Schauer2017}. 
Since they are not ionising photons, they escape the clusters at the same rate at which they are produced (in essence we assume that the H\textsc{ii} photons take care of the recombined atoms and the LW photons escape freely).

\subsection{Escape of ionising photons from Pop~III hosts}
\label{subsec:escape}
Assuming ionisation equilibrium in the ionised medium, the number of escape ionising photons equals the number of ionising photons produced by stars minus the number of recombinations, as recombinations have to be undone by absorbing incoming photons to keep the medium ionised.

\subsubsection{Production: Pop~III cluster model}
\label{subsubsec:stochasticity}
Considering the small number of Pop~III stars formed per halo (especially for low-mass haloes), instead of computing the IMF-averaged escaping photon number for a given halo mass $M_{h}$ and redshift $z$, we do the computation for $N_{\rm cl}=100$ stochastic Pop III clusters and then take the average escape rate at each timestep. In each cluster, we sample stars from a power-law initial mass function (IMF) $dN/dm_{\star}\sim m_\star^{-\alpha}$ between $m_{\rm min}=9~M_\odot$ and $m_{\rm max}=300~M_\odot$, for a total stellar mass $M_\star = \epsilon_\star(\Omega_{\rm b}/\Omega_{\rm m})M_h$\footnote{Throughout our calculations, we adopt the \textit{Planck} cosmological parameters for $\Lambda$CDM, $\Omega_{\rm m}=0.3111$, $\Omega_{\rm b}=0.0490$ and $h=0.6766$ \citep{planck18}.}. The only free parameters here are the IMF power-law index $\alpha$ and star formation efficiency (SFE) $\epsilon_\star$. In total, we explore four scenarios: the `Fiducial' has $\alpha=-1, \epsilon_\star=0.001$; the `HighSFE' has $\alpha=-1, \epsilon_\star=0.002$; the `Salpeter' has $\alpha=-2.35, \epsilon_\star=0.001$; and the `Salpeter \& HighSFE' has $\alpha=-2.35, \epsilon_\star=0.002$. We also generate clusters where the maximum stellar mass is $m_{\rm max}=120~M_\odot$, but the effect, especially on chemically homogeneous models, is very minor. This is because the number of emitted photons per baryon over the stellar lifetime is mostly flat for $m_\star \gtrsim 100~M_\odot$. For this reason we do not show the results for clusters with this smaller $m_{\rm max}$.\\
The sampling method is the following: as long as the total stellar mass $M_{\rm tot} < M_\star$, we keep sampling stars from the IMF, bounded by $m_{\rm min}$ and $m_{\rm max}$. The sampling is finished when we obtain $M_{\star}\le M_{\rm tot}<1.05 M_{\star}$. If we overshoot ($M_{\rm tot} > 1.05 M_\star$), we remove all stars and start again. Then, for each cluster $i \in [1, N_{\rm cl}]$ we can compute the ionising photon production rate in the wavelength range $\Delta\lambda$ at time $t$:
\begin{equation}
\dot{Q}_{\star, i}=\dot{Q}_{\star, i}(\Delta\lambda,t)=\sum_{j=1}^{N_\star(t)}{\dot{N}_{\rm \gamma}(\Delta\lambda,t,m_{\star,j})},
\end{equation}
where $N_\star(t)$ is the total number of stars still alive at time $t$ and ${\dot{N}_{\rm \gamma}(\Delta\lambda,t,m_{\star,j})}$ the production rate of photons in the wavelength range $\Delta\lambda$ at time $t$ for star $j$, given by Eq.~(\ref{eqn8}).

\subsubsection{Absorption: The champagne flow solution}
\label{subsubsec:champagne}

After being emitted by Pop~III stars, ionising photons will be absorbed by the interstellar medium (ISM) of the host halo and only a fraction of them can escape into the IGM. To model this process, we use the idealised semi-analytical model for Pop~III ionisation feedback in \citet{Alvarez2006} under spherical symmetry, which is meant to capture the late-stage overall behaviour of the ISM. Below we summarise the basic ideas of this model and the readers are referred to \citet{Alvarez2006} for details. We note that the propagation of ionising photons in Pop~III star-forming clouds and haloes has been studied with more complex semi-analytical models and simulations \citep[see e.g.][]{Fernandez2011,Paardekooper2015,Tanaka2018,sugimura2020birth,Jaura2022}. We adopt the simple model in \citet{Alvarez2006} to efficiently investigate the parameter space of Pop~III star formation and demonstrate the effects of CHE. In the next section, we compare our results with previous studies to justify our approach. 

For a Pop~III cluster $i$ embedded in a halo of mass $M_h$ at redshift $z$, the ionisation feedback is modelled as a champagne flow bounded by a (D-type)\footnote{The D-type phase of the ionising process (D for Dense) follows the R-type phase (R for rarefied). In the D-phase, the ionisation front moves at subsonic speed relative to the ionised gas and at supersonic speed relative to the neutral gas.} shock running into the primordial star-forming cloud whose initial density profile follows a singular isothermal sphere (SIS)\footnote{It is found in cosmological hydrodynamic simulations that the distribution of primordial gas follows $\rho\propto r^{-2.2}$ in star-forming molecular-cooling minihaloes \citep[e.g.][]{gao2007first,hirano2015primordial} and $\rho\propto r^{-2}$ at $r\gtrsim 0.003\ \rm pc$ in atomic-cooling haloes \citep[e.g.][]{Shang2010,Regan2014,Wise2019}. Therefore, SIS is a good model of the (initial) gas density profile in high-$z$ star-forming haloes.} with a temperature $T_{1}$ \citep[see Eq.~(1) in][]{Alvarez2006}. 
The shock heats the gas up to a temperature $T_{2}\sim 2\times 10^{4}\ \rm K$ such that the high pressure homogenises the downstream medium (i.e. the medium between the star and the inoisation front) and drives an outflow. 
The initial (upstream) temperature $T_{1}$ is related to the (normalisation of) gas density profile $\rho(r)$ assuming that in the central region 
dominated by gas (i.e. ignoring dark matter), gravity is balanced by the thermal pressure of gas\footnote{In our model, $T_{1}$ is an estimate of the gas temperature in the central region ($\lesssim 1\ \rm pc$, where dark matter is relatively unimportant) when the ionisation or shock front breaks out of the central star-forming disk into the cloud. 
At this particular moment, gravity is balanced by pressure at the edge of disk, while thereafter the ionisation or shock front drives an outflow in the cloud. Our model only captures this outflow phase. }. 
Considering the total mass of gas in the halo $4\pi\int_{0}^{R_{\rm vir}}\rho(r|T_{1})r^{2}dr=(\Omega_{\rm b}/\Omega_{\rm m})M_{h}$, we can express $T_{1}$ with halo properties:
\begin{equation}
\left(\frac{T_{1}}{300\ \mathrm{K}}\right)=(\Omega_{\rm b}/\Omega_{\rm m})\left(\frac{M_h}{950\ \mathrm{M_{\odot}}}\right)\left(\frac{R_{\rm vir}}{1\ \mathrm{ pc}}\right)^{-1},
\end{equation}\\
where $R_{\rm vir}\equiv R_{\rm vir}(M_h,z)$ is the virial radius.

Under this setup, the (number) density profile of the gas cloud $n(r,t)\equiv n(r,t|M_h,z)$ is described by the champagne flow solution\footnote{It is shown in \citet[see their Fig.~2]{Tanaka2018} by 1-D radiation
hydrodynamics simulations that the champagne flow solution adequately captures the gas density structure in minihaloes.} in \citet{Shu2002}, governed by the parameter
\begin{equation}
\epsilon=\left(\frac{c_{s,1}}{c_{s,2}}\right)^{2} = \frac{T_1}{T_2}\frac{\mu_2}{\mu_1}\ .
\end{equation}
Here we have $\mu_{2}\simeq 0.63$ for the shocked ionised downstream region, while for the unshocked region, we have $\mu_{1}\simeq 1.22$ (for neutral gas). The quantities $c_{s,1}$ and $c_{s,2}$ are respectively the sound speed in the neutral, ionised region. The densities of the neutral and ionised medium are obtained using Table 1 of \citet{Shu2002} linking values of $\epsilon$ to these densities.

In the early stage, the downstream medium is not fully ionised (due to the high density, see Fig.~\ref{fig:champagne}) and no ionising photons can escape the halo.
Later on, as density decreases by the outflow, the entire downstream is ionised. A part of the ionising photons emitted by the star is used to keep the ionised downstream region ionised, while the rest can ionise the outer region beyond the shock front and even escape the halo.
The rate (luminosity) of escaping ionising photons in the wavelength range $\Delta\lambda$ can be written as:
\begin{align}
\label{eqn12}
\dot{Q}_{\star,{\rm esc}, i}&\equiv \dot{Q}_{\star,{\rm esc}, i}(\Delta\lambda,t,M_h,z)\notag\\
&=\max\left\{\dot{Q}_{\star, i}-4\pi\alpha^{\rm B}_{\rm k}\int_{0}^{10 R_{\rm vir}}r^{2}n_{\rm k}^{2}(r,t)dr, ~0\right\}.
\end{align}
Here the first term on the right hand side is the production rate of ionising photons by all the stars in the cluster, defined in the previous section.
The second term is the recombination rate within 10 times the halo virial radius $R_{\rm vir}$, which is dominated by the downstream medium within the shock front. Here we chose $10 R_{\rm vir}$ as the boundary between the halo and the IGM.
$n_{\rm k}(r,t)\equiv x_{\rm k} n(r,t)$ is the density profile and $\alpha^{\rm B}_{\rm k}$ is the case-B recombination coefficient for species k that corresponds to $\Delta\lambda$. For our three species (H\textsc{ii}, He\textsc{ii}, He\textsc{iii}), the abundances are $x_{\rm H}=0.93$, $x_{\rm He}=0.07$, and we use the formulas of $\alpha^{\rm B}_{\rm k}$ in Appendix E of \citet{Rosdahl2013}, all in units of [cm$^3$ s$^{-1}$]:
\begin{align*}
    \alpha^{\rm B}_{\rm H\textsc{ii}} &= 2.753\times10^{-14} \frac{\lambda_{\rm H\textsc{i}}^{1.5}}{\left[1 + \left(\lambda_{\rm H\textsc{i}}/2.74\right)^{0.407}\right]^{2.242}},\\
    \alpha^{\rm B}_{\rm He\textsc{ii}} &= 1.26\times10^{-14} \lambda_{\rm He\textsc{i}}^{0.75},\\
    \alpha^{\rm B}_{\rm He\textsc{iii}} &= 5.506\times10^{-14} \frac{\lambda_{\rm He\textsc{ii}}^{1.5}}{\left[1 + \left(\lambda_{\rm He\textsc{ii}}/2.74\right)^{0.407}\right]^{2.242}},
\end{align*}
where $\lambda_{\rm H\textsc{i}}= \frac{315614 \rm{K}}{T\rm{[K]}}$, $\lambda_{\rm He\textsc{i}}= \frac{570670 \rm{K}}{T\rm{[K]}}$, and $\lambda_{\rm He\textsc{ii}}= \frac{1263030 \rm{K}}{T\rm{[K]}}$, and $T=T_{2}=2\times 10^{4}\ \rm K$ in our case. 
The whole procedure described above is applied to $N_{\rm cl}$ clusters to obtain the average (over all clusters) luminosity of escaping photons $\dot{Q}_{\star,\rm esc}(\Delta\lambda,t,M_h,z)=\sum_{i=1}^{N_{\rm cl}}\dot{Q}_{\star,{\rm esc},i}/N_{\rm cl}$. At last, the average escaping ionising photon number per halo can be derived as $Q_{\star,\rm esc}(\Delta\lambda,M_h,z)=\int \dot{Q}_{\star,\rm esc}(\Delta\lambda,t,M_h,z) dt$, where we integrate over the cluster lifetime (that is, the lifetime of the longest-lived star in the cluster). The average escape fraction of ionising photons as a function of time is $f_{\rm esc}(\Delta\lambda,t,M_h,z)\equiv \dot{Q}_{\star,\rm esc}(\Delta\lambda,t,M_h,z)/\dot{Q}_{\star}(\Delta\lambda,t,M_h,z)$, where $\dot{Q}_{\star}(\Delta\lambda,t,M_h,z)$ is the average production rate of ionising photons, and the overall escape fraction is $\hat{f}_{\rm esc}(\Delta\lambda,M_h,z)\equiv Q_{\star,\rm esc}(\Delta\lambda,M_h,z)/Q_{\star}(\Delta\lambda,M_h,z)$ given the total number of photons produced $Q_{\star}(\Delta\lambda,M_h,z)=\int \dot{Q}_{\star}(\Delta\lambda,t,M_h,z) dt$. In other words:
\begin{equation}
\label{eqn13}
    \hat{f}_{\rm esc}(\Delta\lambda,M_h,z) = \frac{\int \dot{Q}_{\star\rm{, esc}}(\Delta\lambda,t,M_h,z)~ dt}{\int \dot{Q}_{\star}(\Delta\lambda,t,M_h,z)~ dt}.
\end{equation}
We can then obtain the cosmic escape fraction as a function of redshift:
\begin{equation}
\label{eqn14}
   \overline{f_{\rm esc}}(\Delta\lambda, z) = \frac{\int_{M_1}^{M_2}{\frac{dn_{h}(M_h,z)}{d\ln{M_h}}f_{\rm esc}(\Delta\lambda,M_h, z)~d\ln{M_h}}}
        {\int_{M_1}^{M_2}{\frac{dn_{h}(M_h,z)}{d\ln{M_h}}~d\ln{M_h}}},
\end{equation}
where $n_h(M_h,z)$ is the halo mass function \citep{Tinker2008}, for which we consider haloes in the mass range defined by ${M_{1}=10^{6}~M_{\odot}}$ and $M_{2}=10^{8}~M_{\odot}$ as the typical hosts of Pop~III stars\footnote{The lower bound ($10^6~M_\odot$) is the critical mass for H$_2$ cooling to be efficient, considering the effect of baryon-dark matter streaming motions \citep{Anna2019}. The upper bound ($10^8~M_\odot$) is the typical halo mass for the first galaxies above which haloes are mostly metal enriched and thus do not form Pop~III stars \citep[e.g.][]{Bromm2011}.}.
\subsection{IGM ionisation fraction}
\label{subsec:IGMioni}

The second step is to calculate the evolution of the IGM ionisation fraction $Q_{\rm ion}$ by combining the contribution of all haloes hosting Pop III stars. We define $Q_{\rm ion}$ in the following way (where we decouple singly and doubly ionised helium for simplicity):
\begin{align*}
\begin{cases}
    Q_{\rm ion, H\textsc{ii}} &= \frac{n_{\rm H\textsc{ii}}}{n_{\rm H}},\\
    Q_{\rm ion, He\textsc{ii}} &= \frac{n_{\rm He\textsc{ii}}}{n_{\rm He}},\\
    Q_{\rm ion, He\textsc{iii}} &= \frac{n_{\rm He\textsc{iii}}}{n_{\rm He}}.
\end{cases}
\end{align*}

For species k, the evolution of $Q_{\rm ion,k}$ follows \citep[see e.g. Sec. 3.1 of][]{Hartwig2015}:
\begin{equation}
\label{eqn15}
\frac{dQ_{\rm ion,k}(z)}{dz}=(x_{\rm k}n_{\rm b})^{-1}\frac{dn_{\rm ion,k}(z)}{dz}-\left[\frac{Q_{\rm ion,k}(z)}{t_{\rm rec,k}(z)}\right]\left|\frac{dt}{dz}\right|,
\end{equation}
given the recombination timescale
\begin{equation}
t_{\rm rec,k}(z)=\frac{1}{C_{\rm IGM}(z)\alpha^{\rm B}_{\rm k}x_{\rm k}n_{\rm b}(1+z)^{3}},
\end{equation}
with the case-B recombination coefficient $\alpha^{\rm B}_{\rm k}$ evaluated at a temperature of $T_{\rm IGM}=10^{4}\ \rm K$ for ionised IGM, and
\begin{equation}
\label{eqn17}
\frac{dn_{\rm ion,k}(z)}{dz}=\int_{M_{1}}^{M_{2}}{\frac{dn_{h}(M_h,z)}{dz} Q_{\star,\rm esc}(\Delta\lambda,M_h,z)~dM_h}.
\end{equation}
Here 
$n_{\rm b}\simeq 2\times 10^{-7}\ \rm cm^{-3}$ is the cosmic comoving number density of baryons, $n_h(M_h,z)$ is the halo mass function, 
and $C_{\rm IGM}(z)$ is the redshift-dependent clumping factor, from \citet{Haardt2012}:
\begin{equation}
    C_{\rm IGM}(z) = 1 + 43 ~ z^{-1.71}.
\end{equation}
We solve the differential equation (\ref{eqn14}) from $z=30$ to $z=15$, taking as initial values $Q_{\rm ion,H\textsc{i}}(z=30)=10^{-4}$ and $Q_{\rm ion,He\textsc{ii}}(z=30)=Q_{\rm ion,He\textsc{iii}}(z=30)=0$ \citep[see Fig.~1 in][]{Galli2013}. 
\section{Results}
\label{sec:results}
\subsection{Stellar models}
\begin{table*}[h]
    \caption{Stellar properties of the CHE and standard Pop III models, at the start and end of the MS.}
    \label{tab:stellar_properties}
    \scriptsize
    \begin{threeparttable}
    \begin{tabular}{l|c|c|c|c|c|c|c|c|c}
        \hline\hline
        Mass & $T_{\rm eff}(X_c=0.75 \xrightarrow{} 0)$ & $T_c(X_c=0.75\xrightarrow{}0)$ & $L(X_c=0.75\xrightarrow{}0)$ & $R_\star(X_c=0.75\xrightarrow{}0)$ & $\tau_{\rm MS}$ & $N_{\gamma}(\rm{H\textsc{ii}})^1$ & $N_{\gamma}(\rm{He\textsc{ii}})^1$ & $N_{\gamma}(\rm{He\textsc{iii}})^1$ & $N_{\gamma}(\rm LW)^1$\\
        $M_\odot$ & $\log_{10}{[K]}$ & $\log_{10}{[K]}$ & $\log_{10}{[L_\odot]}$ & $R_\odot$ & Myr & & & &\\
        \hline
        9 CHE & 4.68 $\xrightarrow{}$ 4.86 & 7.86 $\xrightarrow{}$ 7.86 & 3.96 $\xrightarrow{}$ 5.17 & 1.41 $\xrightarrow{}$ 2.46 & 27.15 & $8.08\times10^{62}$ & $3.00\times 10^{62}$ & $4.54\times 10^{60}$ & $2.98\times10^{62}$\\
        9 standard & 4.63 $\xrightarrow{}$ 4.60 & 7.86 $\xrightarrow{}$ 8.08 & 3.72 $\xrightarrow{}$ 4.08 & 1.31 $\xrightarrow{}$ 2.26 & 17.73 & $2.00\times10^{62}$ & $3.20\times 10^{61}$ & $5.14\times10^{58}$ & $1.02\times10^{62}$\\
        12 CHE & 4.71 $\xrightarrow{}$ 4.88 & 7.88 $\xrightarrow{}$ 7.88 & 4.30 $\xrightarrow{}$ 5.39 & 1.76 $\xrightarrow{}$ 2.86 & 17.45 & $1.55\times10^{63}$ & $6.69\times10^{62}$ & $1.32\times10^{61}$ & $5.37\times10^{62}$\\
        12 standard & 4.67 $\xrightarrow{}$ 4.64 & 7.85 $\xrightarrow{}$ 8.11 & 4.11 $\xrightarrow{}$ 4.46 & 1.75 $\xrightarrow{}$ 3.03 & 17.86 & $4.96\times10^{62}$ & $1.11\times10^{62}$ & $4.21\times10^{59}$ & $2.21\times10^{62}$\\
        15 CHE & 4.76 $\xrightarrow{}$ 4.91 & 7.93 $\xrightarrow{}$ 7.93 & 4.56 $\xrightarrow{}$ 5.55 & 1.92 $\xrightarrow{}$ 2.96 & 12.83 & $2.43\times10^{63}$ & $1.26\times10^{63}$ & $3.56\times10^{61}$ & $7.77\times10^{62}$\\
        15 standard & 4.73 $\xrightarrow{}$ 4.66 & 7.93 $\xrightarrow{}$ 8.12 & 4.33 $\xrightarrow{}$ 4.71 & 1.68 $\xrightarrow{}$ 3.63 & 13.00 & $9.09\times10^{62}$ & $2.63\times10^{62}$ & $1.84\times10^{60}$ & $3.65\times10^{62}$\\
        20 CHE & 4.81 $\xrightarrow{}$ 4.95 & 7.99 $\xrightarrow{}$ 7.99 & 4.87 $\xrightarrow{}$ 5.75 & 2.16 $\xrightarrow{}$ 3.11 & 9.07 & $4.10\times10^{63}$ & $2.63\times10^{63}$ & $1.10\times10^{62}$ & $1.20\times10^{63}$\\
        20 standard & 4.79 $\xrightarrow{}$ 4.68 & 7.99 $\xrightarrow{}$ 8.14 & 4.66 $\xrightarrow{}$ 5.04 & 1.83 $\xrightarrow{}$ 4.69 & 9.52 & $1.89\times10^{63}$ & $7.07\times10^{62}$ & $9.25\times10^{60}$ & $6.84\times10^{62}$\\
        30 CHE & 4.87 $\xrightarrow{}$ 4.99 & 8.04 $\xrightarrow{}$ 8.04 & 5.27 $\xrightarrow{}$ 6.00 & 2.66 $\xrightarrow{}$ 3.53 & 6.11 & $8.01\times10^{63}$ & $6.22\times10^{63}$ & $3.81\times10^{62}$ & $2.17\times10^{63}$\\
        30 standard & 4.86 $\xrightarrow{}$ 4.71 & 8.04 $\xrightarrow{}$ 8.17 & 5.09 $\xrightarrow{}$ 5.44 & 2.19 $\xrightarrow{}$ 6.51 & 6.13 & $4.66\times10^{63}$ & $2.28\times10^{63}$ & $5.51\times10^{61}$ & $1.52\times10^{63}$\\
        40 CHE & 4.89 $\xrightarrow{}$ 5.00 & 8.06 $\xrightarrow{}$ 8.06 & 5.52 $\xrightarrow{}$ 6.17 & 3.17 $\xrightarrow{}$ 4.01 & 4.89 & $1.25\times10^{64}$ & $1.06\times10^{64}$ & $7.61\times10^{62}$ & $3.28\times10^{63}$\\
        40 standard & 4.90 $\xrightarrow{}$ 4.73 & 8.06 $\xrightarrow{}$ 8.18 & 5.37 $\xrightarrow{}$ 5.69 & 2.57 $\xrightarrow{}$ 8.17 & 4.78 & $8.23\times10^{63}$ & $4.65\times10^{63}$ & $1.55\times10^{62}$ & $2.55\times10^{63}$\\
        60 CHE & 4.92 $\xrightarrow{}$ 5.02 & 8.08 $\xrightarrow{}$ 8.08 & 5.84 $\xrightarrow{}$ 6.40 & 4.05 $\xrightarrow{}$ 4.86 & 3.82 & $2.25\times10^{64}$ & $2.08\times10^{64}$ & $1.76\times10^{63}$ & $5.72\times10^{63}$\\
        60 standard & 4.94 $\xrightarrow{}$ 4.73 & 8.08 $\xrightarrow{}$ 8.19 & 5.73 $\xrightarrow{}$ 6.00 & 3.24 $\xrightarrow{}$ 11.51 & 3.67 & $1.68\times10^{64}$ & $1.10\times10^{64}$ & $5.23\times10^{62}$ & $4.92\times10^{63}$\\
        85 CHE & 4.94 $\xrightarrow{}$ 5.03 & 8.09 $\xrightarrow{}$ 8.09 & 6.09 $\xrightarrow{}$ 6.59 & 4.98 $\xrightarrow{}$ 5.75 & 3.25 & $3.61\times10^{64}$ & $3.52\times10^{64}$ & $3.32\times10^{63}$ & $8.96\times10^{63}$\\
        85 standard & 4.96 $\xrightarrow{}$ 4.73 & 8.09 $\xrightarrow{}$ 8.20 & 6.00 $\xrightarrow{}$ 6.23 & 3.97 $\xrightarrow{}$ 15.20 & 3.08 & $2.90\times10^{64}$ & $2.09\times10^{64}$ & $1.23\times10^{63}$ & $8.22\times10^{63}$\\
        120 CHE & 4.95 $\xrightarrow{}$ 5.03 & 8.10 $\xrightarrow{}$ 8.10 & 6.32 $\xrightarrow{}$ 6.77 & 6.09 $\xrightarrow{}$ 6.84 & 2.86 & $5.63\times10^{64}$ & $5.72\times10^{64}$ & $5.81\times10^{63}$ & $1.38\times10^{64}$\\
        120 standard & 4.98 $\xrightarrow{}$ 4.72 & 8.10 $\xrightarrow{}$ 8.21 & 6.25 $\xrightarrow{}$ 6.44 & 4.89 $\xrightarrow{}$ 20.27 & 2.74 & $4.78\times10^{64}$ & $3.69\times10^{64}$ & $2.53\times10^{63}$ & $1.33\times10^{64}$\\
        180 CHE & 4.95 $\xrightarrow{}$ 5.03 & 8.09 $\xrightarrow{}$ 8.09 & 6.57 $\xrightarrow{}$ 6.97 & 7.91 $\xrightarrow{}$ 8.66 & 2.54 & $9.48\times10^{64}$ & $9.72\times10^{64}$ & $1.00\times10^{64}$ & $2.31\times10^{64}$\\
        180 standard & 4.99 $\xrightarrow{}$ 4.52 & 8.09 $\xrightarrow{}$ 8.23 & 6.51 $\xrightarrow{}$ 6.68 & 6.40 $\xrightarrow{}$ 65.09 & 2.88 & $8.35\times10^{64}$ & $6.57\times10^{64}$ & $5.15\times10^{63}$ & $2.41\times10^{64}$\\
        250 CHE & 4.96 $\xrightarrow{}$ 5.04 & 8.09 $\xrightarrow{}$ 8.09 & 6.75 $\xrightarrow{}$ 7.13 & 9.60 $\xrightarrow{}$ 10.35 & 2.36 & $1.41\times10^{65}$ & $1.47\times10^{65}$ & $1.55\times10^{64}$ & $3.42\times10^{64}$\\
        250 standard & 4.99 $\xrightarrow{}$ 4.33 & 8.09 $\xrightarrow{}$ 8.23 & 6.72 $\xrightarrow{}$ 6.86 & 7.92 $\xrightarrow{}$ 195.50 & 3.08 & $1.27\times10^{65}$ & $1.02\times10^{65}$ & $8.51\times10^{63}$ & $3.71\times10^{64}$\\
        300 CHE & 4.96 $\xrightarrow{}$ 5.04 & 8.09 $\xrightarrow{}$ 8.09 & 6.85 $\xrightarrow{}$ 7.22 & 10.54 $\xrightarrow{}$ 11.28 & 2.27 & $1.74\times10^{65}$ & $1.83\times10^{65}$ & $2.00\times10^{64}$ & $4.19\times10^{64}$\\
        300 standard & 4.99 $\xrightarrow{}$ 4.01 & 8.09 $\xrightarrow{}$ 8.23 & 6.82 $\xrightarrow{}$ 6.95 & 8.80 $\xrightarrow{}$ 905.61 & 3.01 & $1.57\times10^{65}$ & $1.28\times10^{65}$ & $1.10\times10^{64}$ & $4.63\times10^{64}$\\
        \hline
    \end{tabular}
    \begin{tablenotes}
    \item[1] $N_\gamma(\Delta\lambda)$ refers to the number of photons emitted in the wavelength range $\Delta\lambda$, integrated over the whole (MS) lifetime of the considered star: ${N_\gamma(\Delta\lambda, m_\star)=\int{\dot{N}_\gamma(\Delta\lambda, t,m_\star)~dt}}$.
    \end{tablenotes}
    \end{threeparttable}
\end{table*}
\begin{figure}[h]
    \centering
    \includegraphics{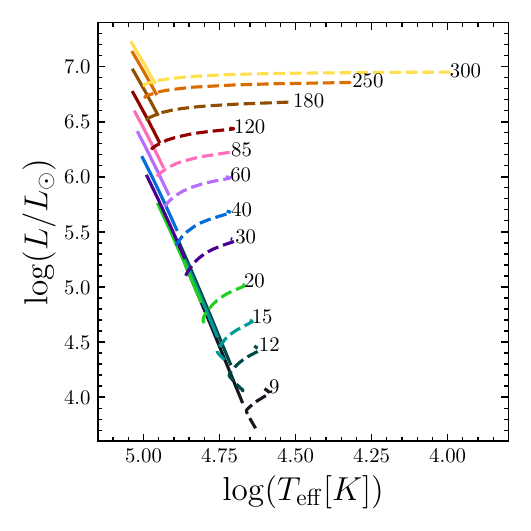}
    \caption{HR diagram showing the MS evolution of the Pop III chemically homogeneous models (solid lines) and that of the Pop III standard models computed with GENEC by \citet{Murphy2021} (dashed lines). The numbers indicate the initial mass in $M_\odot$ of each model.}
    \label{fig:HRD}
\end{figure}
We compare the evolution in the HR diagram, lifetimes of the MS, as well as photon production over time, for stars computed with our analytical Pop III CHE model and standard Pop III stars computed with GENEC by \citet{Murphy2021}. Figure~\ref{fig:HRD} shows the evolution in the HR diagram of the CHE (solid) and standard (dashed) Pop III models during their MS. Right at the onset of hydrogen burning the standard models have a chemically homogeneous composition, so we would expect the starting position of CHE and classical models of the same mass to be at the same position in the HR diagram. However, there are small differences between the CHE and classical models, which come from the fact that polytropes are only providing an approximation of the structure of a star. Let us remind that using a polytropic model implicitely assumes that the ratio of the gas to the total pressure $\beta$ is uniform in the star, which is not strictly the case, although $\beta$ does not vary much over most of the mass of the star.

The most striking feature of the CHE models is their blue and upwards evolution in the HRD. Indeed, the very strong mixing characteristic of CHE models removes the core-envelope differentiation. This means that the traditional hydrogen envelope expansion that occurs during the MS does not happen for those models. The chemically homogeneous models do expand slightly during this phase; both their luminosity and effective temperatures increase, but $T_{\rm eff}$ does not grow fast enough to keep the radius constant. That expansion however is limited to a factor of 1.7 between the initial and final radii, whereas for the standard models it can reach a factor of 100. 
Interestingly, this factor decreases with mass for the CHE models (the final to initial radius ratio being 1.7 for the 9 $M_\odot$ star and reaching 1.07 for the 300 $M_\odot$ star) whereas it increases with mass for the classical models (the final to initial radius ratio also being 1.7 for the 9 $M_\odot$ star, and reaching 100 for the 300 $M_\odot$ star).\\
Apart for the 9 $M_\odot$ and 12 $M_\odot$ stars where the CHE models have much longer MS lifetimes, both sets of models exhibit very similar MS durations. This can be explained by the competition of two effects. On the one hand, the CHE models have a much larger fuel supply since the entire initial mass of hydrogen will burn into helium: this should increase the lifetime of the MS. On the other hand, the CHE models are hotter and more luminous than the classical models, which should decrease the lifetime of the MS. The result, as stated above, is that masses above 12 $M_\odot$ do not exhibit significant differences in their MS duration. The difference in lifetimes between the lower mass stellar models can be seen in the late evolution of the clusters, where no more photons are emitted after $\sim 18$ Myr in the clusters using standard stars whereas the clusters with chemically homogeneous stars last until $\sim 27$ Myr. All these results, as well as the numbers of photons emitted by the individual stellar models over their MS lifetimes in our different wavelength ranges of interest, are summarised in Table~\ref{tab:stellar_properties}.
\begin{figure}[h]
    \centering
    \includegraphics{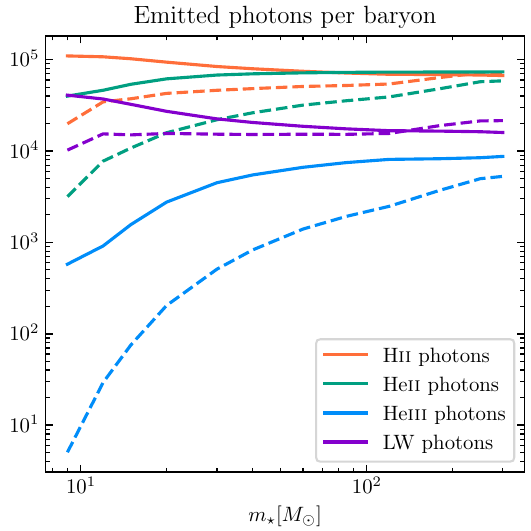}
    \caption{Number of photons emitted per baryon over the MS lifetime, for each stellar mass and for the four wavelength ranges considered (H\textsc{ii} photons: $13.6 \rm{eV} < E < 24.6 \rm{eV}$, He\textsc{ii} photons: $24.6 \rm{eV} < E < 54.4 \rm{eV}$, He\textsc{iii} photons: $\rm E > 54.4\rm{eV}$, LW photons: $11.2\rm{eV} < E < 13.6\rm{eV}$). Instead of plotting individual data points corresponding to individual masses we join them for better clarity. The solid lines represent the CHE models and the dashed ones, the classical models computed with GENEC.}
    \label{fig:ppb}
\end{figure}

Figure~\ref{fig:ppb} shows the number of photons emitted per baryon integrated over the MS lifetime as a function of stellar mass, for the four wavelength ranges considered. This number incorporates both the photon emission rates and the duration of the MS. Comparing photon production, the CHE models emit systematically more photons than their classical counterparts, except for the LW band. This makes sense as the CHE stars are more luminous (more photons emitted in total), and have higher effective temperatures (more energetic photons emitted). For the LW photons, their wavelengths correspond to black-body temperatures of $26 - 32 \times 10^4 \rm{K}$ (or $\log_{10}(T_{\rm eff}\rm{[K]}) \sim 4.4 - 4.5$), which are closer to the effective temperatures of the standard models than to those of the CHE models. As such, the effect of chemically homogeneous evolution on the LW feedback should be minor. Still, the production of ionising photons (whether H ionising or He singly or doubly ionising) is higher for the CHE than the standard models by a factor of $\sim 1.5 - 8$ (decreasing with stellar mass).\\
For the CHE models, it is notable that the rate of emitted H\textsc{ii} and He\textsc{ii} photons per baryon does not depend on stellar mass for $ m_\star\gtrsim 100~M_\odot$. The overall values are consistent in an order of magnitude with those in Fig.~1 of \citet{Heger2010}.

\subsection{Escaping photons}

\begin{figure}[h!]
    \centering
    \includegraphics{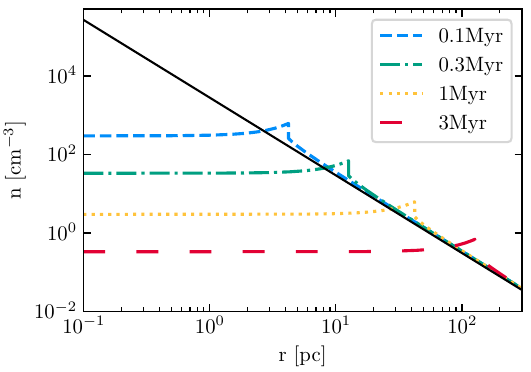}
    \caption{Density profiles obtained from the champagne flow solution \citep{Shu2002} at $t$ = 0.1 (dashed), 0.3 (dash-dotted), 1 (dotted), and 3 Myr (long-dashed), on top of the initial density profile (solid) for a typical primordial star-forming cloud in a DM minihalo with $M_h=10^6~M_\odot$ at $z=20$. 
    }
    \label{fig:champagne}
\end{figure}

In the above subsection, we looked at the photons emitted by stars, without accounting for the surrounding environment. Here, we focus on the effect of the absorption due to recombinations in the halo, corresponding to the second term on the right-hand side of Eq.~(\ref{eqn12}), which integrates the density profile of the champagne flow solution at each time. For illustration purposes, we show in Fig.~\ref{fig:champagne} such density profiles at $t$ = 0.1, 0.3, 1, and 3 Myr, for a typical Pop III halo with $M_h=10^6~M_\odot$ at $z=20$.

We can now take into account the recombinations happening both downstream and upstream of the shock front, removing stars that have reached the end of their MS lifetime, at every timestep for the CHE and the classical Pop III models.
\begin{figure*}
    \centering
    \includegraphics[scale=1]{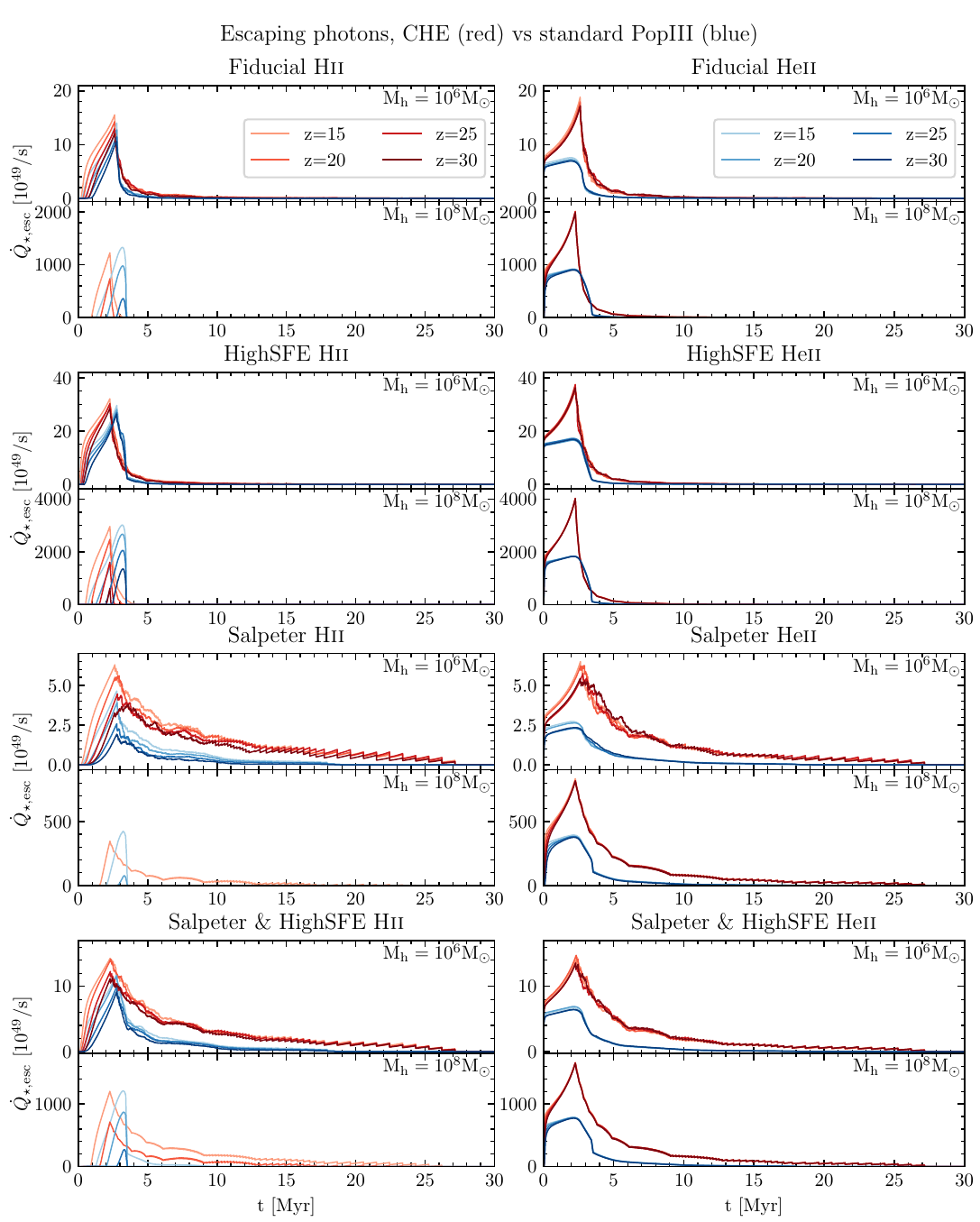}
    \caption{Number of escaping H\textsc{ii} and He\textsc{ii} photons per second ($\dot{Q}_{\star,\rm esc}(\Delta\lambda,t,M_h,z)$) as a function of time, for clusters in a minihalo ($M_h=10^6~M_\odot$) and clusters in an atomic-cooling halo ($M_h=10^8\ M_\odot$). At each integer redshift value between 30 and 15 we generate 100 stochastic clusters and show the average number of escaping photons from these clusters at each timestep. Each of the four rows represents a different set of model parameters (see Table \ref{tab:cluster_scenarios}).}
    \label{fig:escape_HII_HeII}
\end{figure*}
In Fig.~\ref{fig:escape_HII_HeII}, we show the number of escaping photons per second in the hydrogen ionisation (left column) and helium first ionisation (right column) wavelength ranges, for a dark matter minihalo of mass $M_h=10^6~M_\odot$ (upper panel of each subplot; CHE models in red, standard models in blue) and an atomic-cooling halo of mass $M_h=10^8~M_\odot$ (lower panel of each subplot, same colours), at redshifts $z=30, 25, 20, 15$. The four cluster scenarios' parameters are given in Table \ref{tab:cluster_scenarios}. 
We recall that the free parameters of the clusters are the IMF slope and star formation efficiency. The panels' vertical axes for haloes of the same mass in the same scenario use the same scale, for easier comparison between H\textsc{ii} and He\textsc{ii} photon escape.\\

\begin{table}[h]
    \caption{Cluster scenarios.}
    \label{tab:cluster_scenarios}
    \centering
    \begin{tabular}{l|c|c}
        \hline\hline
        Name & IMF slope & $\epsilon_\star$\\
        \hline
        Fiducial & -1 & 0.001 \\
        HighSFE & -1 & 0.002 \\
        Salpeter & -2.35 & 0.001\\
        Salpeter \& HighSFE & -2.35 & 0.002\\
        \hline
    \end{tabular}
\end{table}

The difference in the escape rate of H\textsc{ii} photons between the CHE and standard models can be interpreted with the higher emission rates of ionising photons and shorter lifetimes of Pop~III stars under CHE compared with those under the standard stellar evolution. The former effect is more significant for lower-mass stars while the latter is only important for massive stars ($m_{\star}>120\ \rm M_{\odot}$). In low-mass haloes ($M_{\rm h}= 10^{6}\ \rm M_{\odot}$), where escape of ionising photons is relatively easy, the first effect (higher emission rates) dominates so that the total number of escaping H\textsc{ii} photons is larger for the CHE models. However, for massive haloes ($M_{\rm h}= 10^{8}\ \rm M_{\odot}$), particularly under the top-heavy IMF (Fiducial and HighSFE), the second effect (shorter lifetimes) becomes dominant, such that the peak of escape rate is reached a few Myr earlier for the CHE models than for the standard ones. Since at later stages, the gas density in the halo (and consequently the rate of recombinations) has decreased, more photons escape in the standard case than with CHE.

The effect of the IMF slope is very striking as the bottom-heavy scenarios yield rates of escaping photons that are much flatter over time than the top-heavy IMFs. Indeed, high mass stars produce more ionising photons over shorter lifetimes than low-mass ones, so the scenarios which create them in larger quantities lead to a more pronounced peak of $\dot{Q}_{\star,\rm{ esc, H\textsc{ii}}}$ around 3 Myr (the MS lifetime of those high mass stars). Interestingly, CHE clusters with the bottom-heavy IMFs have more escaping photons when integrating over their whole lifetime than the top-heavy ones. This is because lower-mass stars have a longer lifetime and keep emitting photons at late cluster times when there are very few recombinations. 

The star formation efficiency $\epsilon_\star$ also has a strong influence on the number of escaping H$\textsc{ii}$ photons, which is understandable since the HighSFE scenarios double the total stellar mass in the clusters but keep the number of recombinations (linked only to the halo mass and redshift) constant.

\begin{figure}[h!]
    \centering
    \includegraphics{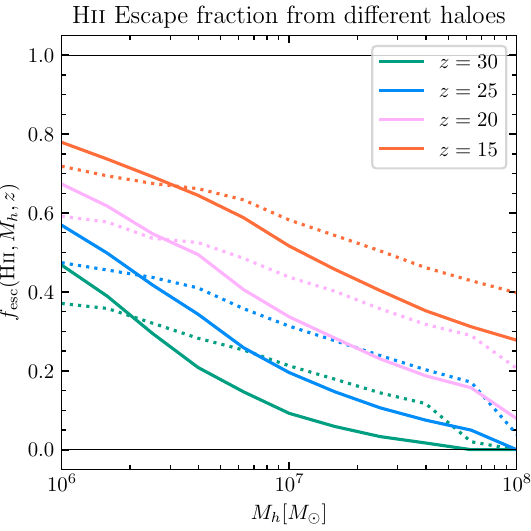}
    \caption{Escape fractions for H\textsc{ii} photons averaged over the lifetimes of the clusters and over the $N_{\rm cl}$ stochastic clusters, $\hat{f}_{\rm esc}(\rm{H\textsc{ii}})$, from fiducial haloes of different masses at four different redshifts. The solid lines show the escape fractions from haloes with chemically homogeneous stars, while the dotted lines show the escape fractions from haloes with classical stars.}
    \label{fig:f_esc_Mh}
\end{figure}

The effect of halo mass, all other parameters being equal, should be to both increase the number of recombinations for higher halo masses, while also increasing the total production of photons (because those haloes produce more stars). It turns out that the effects do not compensate each other: there are about 100 times as many ionising photons produced by the haloes with $M_h=10^8~M_\odot$ than by those with $M_h=10^6~M_\odot$. However, the escape fractions are much lower for the high mass haloes than for the low-mass ones (see Fig.~\ref{fig:f_esc_Mh}).

Finally, for all sets of parameters, we find that there are more escaping H$\textsc{ii}$ photons at the lower redshifts than at higher ones. This is because redshift in our model does not influence how many photons are produced, but only affects the number of recombinations in the halo. Indeed, for the same halo mass, its virial radius $R_{\rm vir}$ increases with time (or with decreasing redshift),
thus the initial upstream temperature $T_1$ decreases with time (or with decreasing redshift)
and so is the \citet{Shu2002} parameter $\epsilon$. As $\epsilon$ decreases towards low redshifts, so do the gas density and the number of recombinations, increasing the number of escaping photons. 


The results above mostly hold for the He\textsc{ii} wavelength range, with the relative difference between CHE and classical models being larger for the escaping He$\textsc{ii}$ photons than for the H$\textsc{ii}$ photons. This is logical considering the black-body temperatures that correspond to He$\textsc{ii}$ photons overlap even more with the effective temperatures of the CHE models in comparison with the standard models, than those corresponding to H$\textsc{ii}$ photons. We note also that halo mass and redshift are less impactful on the escape rate of He$\textsc{ii}$ photons relative to H$\textsc{ii}$ photons. That is because the number of recombining He nuclei is very small in all cases (leading to escape fractions close to 1 for all four scenarios and both types of stellar models).

We do not show the equivalent to Fig.~\ref{fig:escape_HII_HeII} for LW and He\textsc{iii} photons as they are relatively less important, but we present the relevant results here. For the Fiducial and HighSFE scenarios the LW emissions are similar between classical and CHE models, whereas the CHE models dominate in the bottom-heavy Salpeter and Salpeter \& HighSFE scenarios. This is linked to what is shown in Fig.~\ref{fig:ppb} (purple curve). Indeed, clusters containing more lower mass stars (the Salpeter and Salpeter \& HighSFE scenarios) emit more LW photons when they have CHE stars than standard ones. Conversely, the standard clusters with a top-heavy IMF actually exhibit a higher maximum number of escaping LW photons.

In the He\textsc{iii} photons wavelength range, the classical Pop III models produce very few photons (not even enough to counterbalance the recombinations in some cases, leading to an escape rate of 0). The effects of redshift, halo mass, and cluster scenario are qualitatively the same as for other wavelength ranges discussed previously. The main features of this wavelength range are that the relative difference in escape rates between the CHE and standard clusters is the largest among all wavelength ranges, and that those escape rates are in any case very low ($\lesssim 10^{49} s^{-1}$ for $10^6~M_\odot$ haloes, $\lesssim 100\times10^{49} s^{-1}$ for $10^8~M_\odot$ haloes).\\

We perform a check of our ionising photons escape model by using version C17.02 of the photoionisation code Cloudy \citep{Ferland2017} to compute the output spectra of a few synthetic clusters embedded in haloes of different masses at different redshifts. The sizes and densities of the haloes are given by the \citet{Shu2002} model. For the Cloudy modelling we assume spherical geometry, constant densities, and primordial chemical abundances, then run the code to convergence. The quantity we compare is the transmitted spectrum, and we do not take into account the diffuse spectrum. We expect significant differences in escaping photon fluxes as our ionisation model is highly idealised. To be specific, the analytical estimations were calculated in strict energy bins (see Sec.~\ref{subsubsec:ionimodel}), and these bins do not take into account the fact that at energies above 24.6 eV, the ionisation cross-sections of hydrogen per hydrogen nucleus are comparable (or sometimes higher) than the helium ones \citep[see Fig.~2 of][]{Glatzle2019}, leading to an underestimation of the amount of hydrogen ionising photons when compared to the Cloudy models. Furthermore, the analytical model does consider the case in which multiple photons with individually too low energies could together ionise atoms (e.g. two LW photons of the right energies could excite then ionise a hydrogen atom). 

However, the differences turn out to be modest, thus we conclude that our simple model is appropriate. In particular, for clusters with high photon production rates (typically the HighSFE scenario, or Fiducial clusters that drew very massive stars), the escape fractions computed by Cloudy are very similar ($\lesssim 2\%$ relative difference). For clusters producing relatively few photons, we underestimate by $\lesssim 10\%$ the H\textsc{ii} photons' escape fraction, and overestimate by $\lesssim 20\%$ the He\textsc{ii} photons' escape fraction. This leads to similar relative differences in the IGM ionisation rates if we consider a worst case scenario where we rescale all the photon escape rates from all clusters by these amounts (even those clusters where the difference between our simple model and the Cloudy output is very small).
In any case, the variation in escape rates between our model and the Cloudy outputs would apply similarly to clusters with CHE stars and to those with classical stars, so the difference in ionising power would still hold.\\

Figure~\ref{fig:f_esc_Mh} shows the variation of the escape fraction of H\textsc{ii} photons as a function of halo mass, at redshifts $z$=30, 25, 20, and 15, for the Fiducial scenario. The CHE clusters are represented by solid lines while dotted lines are for the clusters containing classical Pop III stars. Interestingly, while the CHE haloes see more escaping photons at most halo masses and redshifts, their escape fractions for the large halo masses are smaller than those of the standard clusters. This is caused by a rather fine effect: the standard stars reach their maximum ionising power slightly later than the CHE models, so the gas density from \citet{Shu2002} and thus the number of recombinations are smaller for the classical haloes. This effect is particularly noticeable in haloes with large numbers of high mass stars (high mass haloes) and recombinations (high mass haloes at high redshift). Overall, we find the expected result that escape fractions decrease with increasing halo mass \citep[as also found by e.g.][]{Wise2014}, and with increasing redshift. Indeed, halo mass scales positively with the number of recombinations, so a larger part of the emitted photons in high mass haloes must be used to reionise the recombined nuclei. On the other hand, recombination rate scales positively with matter density, so that there are fewer recombinations at a later time (lower redshift), leading to increased escape fractions.

We draw attention to the fact that the escape fractions we show in Fig.~\ref{fig:f_esc_Mh} are averaged over the lifetimes of the clusters, as well as over the 100 stochastic clusters. Therefore we do not show the time or the stochastic variability of the escape fractions. The time variability is quite extreme, as in a given cluster the escape fraction can take values from 0 to $\sim 0.9$. On the other hand, the stochastic variability in the values of $\hat{f}_{\rm esc}$ (see Eq.~(\ref{eqn13}) is milder, with relative differences in lifetime-integrated escape fractions of $\sim 10\%$ between the maximally and minimally escaping clusters.

\subsection{IGM ionisation}
\begin{figure}[h!]
    \centering
    \includegraphics{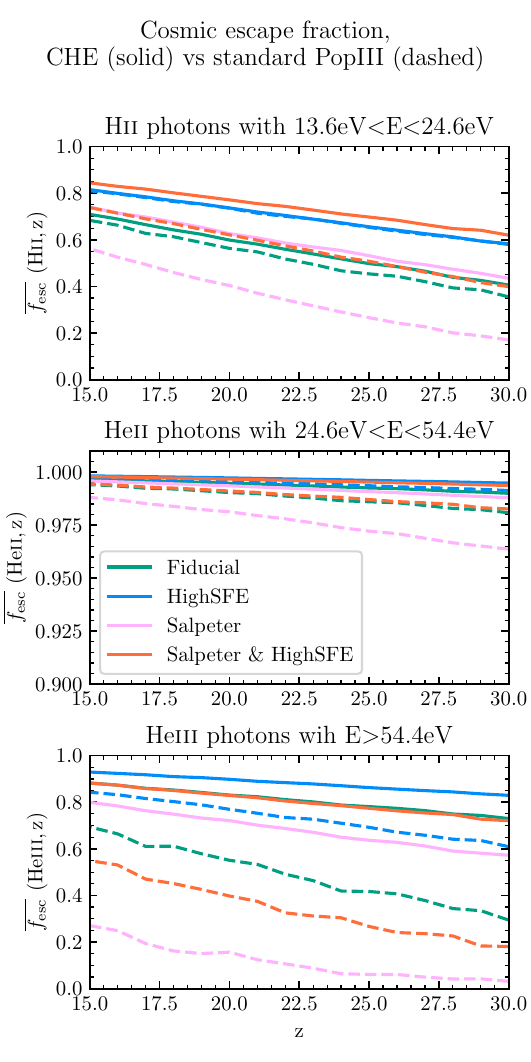}
    \caption{Cosmic escape fraction $\overline{f_{\rm esc}}(\Delta\lambda, z)$ (given by Eq.~\ref{eqn14}). \textit{Top panel}: H\textsc{ii} photons ($13.6 \rm{eV} < E < 24.6 \rm{eV}$), \textit{middle panel}: He\textsc{ii} photons ($24.6 \rm{eV} < E < 54.4 \rm{eV}$), \textit{bottom panel}: He\textsc{iii} photons ($\rm E > 54.4\rm{eV}$). The \textit{middle panel} (He\textsc{ii} photons) is zoomed-in between 0.9 and 1 to better distinguish the different curves.}
    \label{fig:f_esc_cosmo}
\end{figure}
We can now convolve the results from stochastic clusters embedded in haloes of specific masses at specific redshifts, with the distribution of halo masses at each redshift (the halo mass function). Figure~\ref{fig:f_esc_cosmo} shows the cosmic escape fraction $\overline{f_{\rm esc}}$ (defined in Eq.~(\ref{eqn14})) as a function of redshift, for the wavelengths corresponding to H\textsc{ii}, He\textsc{ii}, and He\textsc{iii} photons (LW photons are not shown as their escape fraction is always 1 in our model). The vertical axis for He\textsc{ii} photons goes from 0.9 to 1 so as to better see the differences between scenarios. For all three wavelength ranges we find that, as expected, the CHE models yield larger cosmic escape fractions than the standard models. Furthermore, almost all He\textsc{ii} photons produced escape their host haloes (at least 95\% of them in the worst case), regardless of the scenarios and stellar models.\\
A result that at first glance may seem counter-intuitive is that the H\textsc{ii} escape fraction is higher for the clusters computed with a Salpeter IMF than for those with a fiducial IMF with $\alpha=-1$ (regardless of the SFE). Indeed, from Fig.~\ref{fig:escape_HII_HeII} we may get the impression that the Fiducial scenario leads to more escaping H\textsc{ii} photons than the Salpeter one. But when we integrate over the lifetime of the clusters, there are more H\textsc{ii} photons escaping from Salpeter clusters. To understand this result, one has to combine two facts here: the Salpeter IMF yields more lower-mass and thus longer-lived stars, and the number of recombinations in a halo decreases over time as the density of the champagne flow solution decreases. As a result the late phases ($t \gtrsim 10$Myr) of the Salpeter clusters contribute a non-negligible fraction of the total number of escaping H\textsc{ii} photons, while in the fiducial clusters most stars will have died by then. This result holds for the chemically homogeneous clusters, but not those comprised of classical Pop~III stars from the GENEC models.
\begin{figure}[h!]
    \centering
    \includegraphics{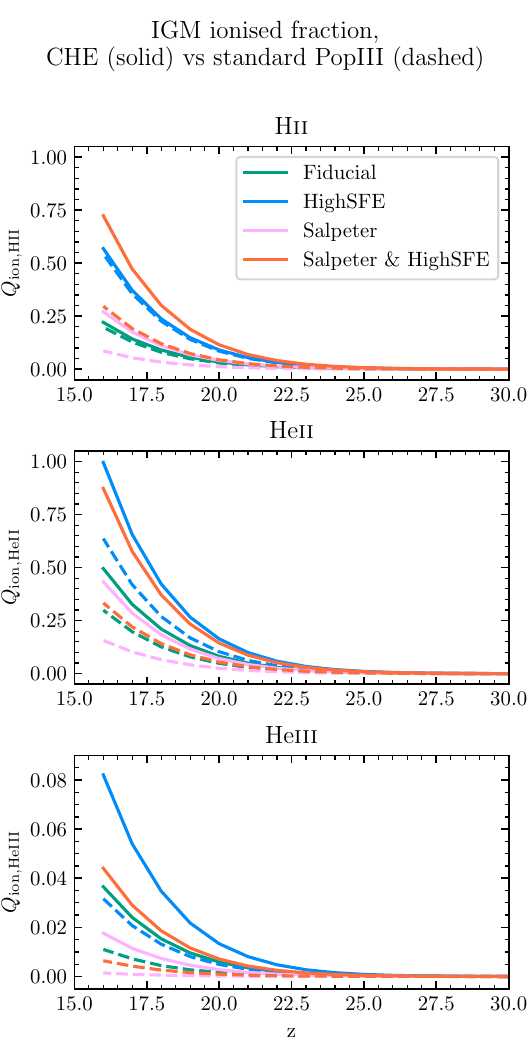}
    \caption{Fraction of ionised species as a function of redshift $Q_{\rm ion, k}(z)$. The \textit{bottom panel} (He\textsc{iii} ionisation fraction) is zoomed-in between 0 and 0.1 to better distinguish the different curves.}
    \label{fig:Qion}
\end{figure}

Figure~\ref{fig:Qion} shows the ionisation fractions of H and He (singly and doubly ionised), $Q_{\rm ion, H\textsc{ii}}$, $Q_{\rm ion, He\textsc{ii}}$, and $Q_{\rm ion, He\textsc{iii}}$. We can see that the standard Pop III scenarios (using GENEC models with fiducial star formation efficiency and either IMF) lead to $Q_{\rm ion, H\textsc{ii}} \sim 0.1-0.2$ at $z\sim 15$. Keeping the same cluster parameters but switching to CHE stellar models increase $Q_{\rm ion, H\textsc{ii}}$ to $\sim 0.2 - 0.3$. Doubling the SFE more than doubles the hydrogen ionisation fraction. This holds true for helium ionisation. As a result, this star formation efficiency parameter $\epsilon_\star$ is a very important one to estimate for Pop~III stars since it strongly influences the impact of these stars on the reionisation. We find slightly higher ionisation fractions for He\textsc{ii} than for H\textsc{ii}, but a very small doubly ionised helium fraction. 

\begin{figure}
    \centering
    \includegraphics[width=1\columnwidth]{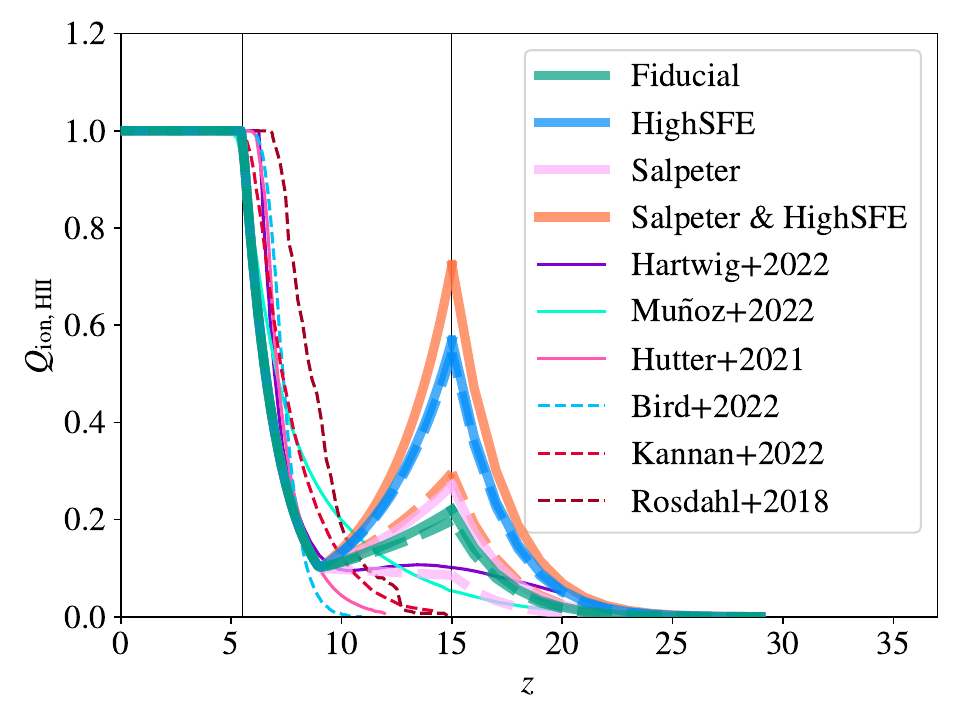}
    \caption{Ionised fraction of hydrogen. Extrapolation of the predictions (at $z\sim 15-30$) from our CHE and standard models are shown with thick solid and dashed curves. For comparison, we plot the results from the semi-analytical models ASLOTH \citep{Hartwig2022}, 21cmFAST \citep{Munoz2022} and Astraeus \citep{Hutter2021}, as well as cosmological simulations ASTRID \citep{Bird2022}, THESAN \citep{Kannan2022} and SPHINX \citep{Rosdahl2018} with thin solid and dashed curves, respectively. The vertical lines show the redshift range ($z\sim 5.5-15$) in which our predictions at high-$z$ are extrapolated to the post-reionisation epoch.}
    \label{qion_comp}
\end{figure}

We then compare our results with previous theoretical predictions and observational constraints on the ionisation history (of hydrogen). We only focus on Pop III stars at $z\sim 15-30$ and do not model the transition between Pop III and metal-enriched Pop~II star formation, which is expected to happen in the same redshift range \citep[$z\sim 15-30$, see e.g.][]{Maio2010,johnson2013first,xu2016late,johnson2013first,Mebane2018,jaacks2018baseline,sarmento2018following,liu2020did}. So we extrapolate the ionised fraction $Q_{\rm ion,HII}$ from our model to $z=0$ with a conservative estimate of the ionisation history in the Pop~II-dominated era ($z\lesssim 15$). We assume that the ionised fraction is 0.1 at $z=9$, appropriate for late reionisation scenarios \citep[e.g.][]{Greif2006}, and increases with decreasing $z$ following a power-law to reach unity at $z=5.5$. Between $z=9$ and $z=15$, we connect the high-$z$ (predicted) and low-$z$ (assumed) pieces with a power-law\footnote{Considering that Pop~II stars are less powerful sources of ionising photons than Pop III stars, it is likely that the ionised fraction is not a monotonic function of $z$, especially when the transition between Pop III and Pop~II is not smooth. We therefore allow $Q_{\rm ion,HII}$ to decrease with decreasing redshift at $z\sim 9-15$ \citep[see also][]{Cen2003}.}. The results are shown in Fig.~\ref{qion_comp}, compared with the predictions from the semi-analytical models ASLOTH \citep{Hartwig2022}, 21cmFAST \citep{Munoz2022} and Astraeus \citep{Hutter2021}, as well as cosmological simulations ASTRID \citep{Bird2022}, THESAN \citep{Kannan2022} and SPHINX \citep{Rosdahl2018}. We find that with the fiducial SFE and Salpeter IMF, the GENEC model is generally consistent with the semi-analytical models that cover $z\gtrsim 15$ \citep[e.g.][]{Hartwig2022,Munoz2022}), while $Q_{\rm ion,HII}$ is a factor of $\sim 2-4$ higher with CHE. With a higher SFE and/or a more top heavy IMF, $Q_{\rm ion,HII}$ is significantly higher for both CHE and GENEC by up to a factor of $\sim 10$. 

\begin{figure}
    \centering
    \includegraphics[width=1\columnwidth]{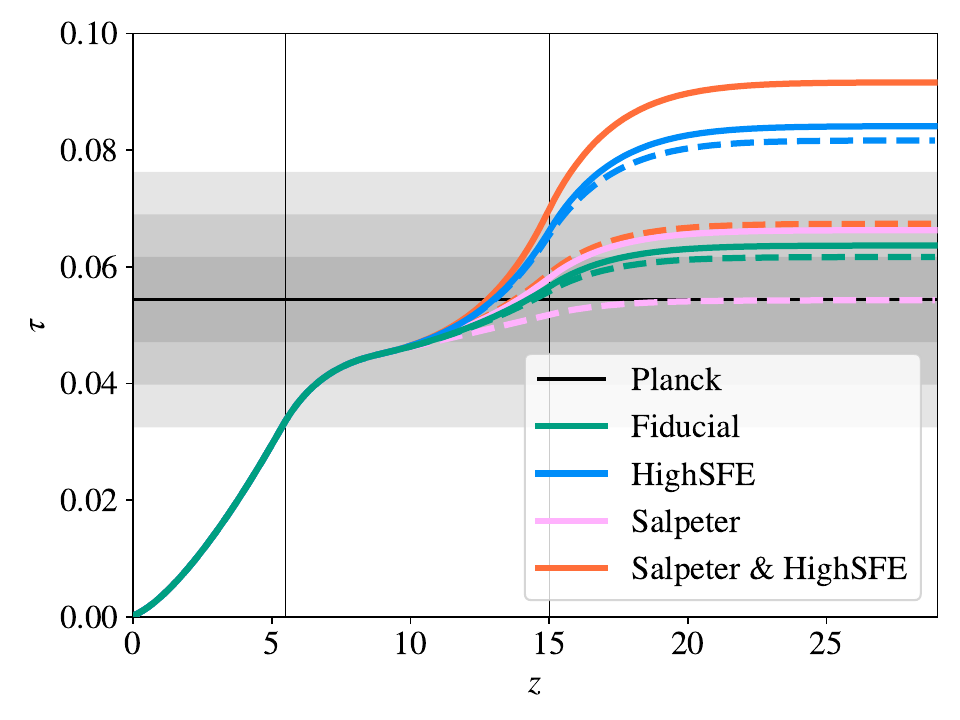}
    \caption{Optical depth to Thomson scattering, based on the extrapolation of the predicted ionisation histories from our CHE and GENEC models (see Fig.~\ref{qion_comp}). The recent measurement by \textit{Planck} \citep{aghanim2020planck} is shown with the horizontal line, whose 1, 2, and 3\ $\sigma$ scatters are denoted by the shaded regions. Again, the vertical lines show the redshift range ($z\sim 5.5-15$) in which our predictions at high-$z$ are extrapolated to the post-reionisation epoch.}
    \label{tau}
\end{figure}

We also calculate the optical depth to Thomson scattering, $\tau$, from the ionisation history, as shown in Fig.~\ref{tau} on top of the value measured by \textit{Planck} $\tau=0.0544\pm 0.0073$ \citep{aghanim2020planck}. The standard GENEC model with the fiducial SFE and Salpeter IMF well reproduces the observed $\tau$. Under the same conditions, the CHE model slightly overpredicts $\tau$ by $\sim 1\sigma=0.0073$. $\tau$ is significantly higher than the observed value with a higher SFE and/or a more top-heavy IMF, by $\sim 1-1.5\sigma$ for the Fiducial GENEC and CHE models, and more than $3\sigma$ in the HighSFE case of CHE with either IMF and GENEC with the fiducial IMF. While in the intermediate cases with fiducial SFE for CHE and enhanced SFE for GENEC, under the Salpeter IMF, $\tau$ is overpredicted by $\sim 1.5-2\sigma$. Our results indicate that the ionisation history and contribution to $\tau$ at $z\gtrsim 15$ (when Pop III stars can play an important role) is highly sensitive to the SFE and IMF. This also implies that the maximum SFE allowed so that the Universe is not ionised too fast is lower if most Pop III stars experience CHE; this effect is more significant for a less top-heavy IMF. Finally, we would like to point out that the above analysis is rather qualitative as $\tau$ is a measurement of the entire ionisation history and our model only covers the high-$z$ regime and relies on extrapolation at low $z$. Beyond $\tau$, the radiation fields from Pop III stars (e.g. ionising, Lyman-$\alpha$, and LW photons) can be better constrained with 21-cm cosmology \citep[see e.g.][]{Fialkov2014,Barkana2016,Cohen2017,Tanaka2018,Mirocha2018,Schauer2019,Mebane2020}. We plan to investigate the impact of CHE on the 21-cm signal in future work.

\section{Discussion}
\label{sec:discussion}

The idea of this work is to remain simple in providing an upper limit to the contribution of single Pop~III stars to cosmic reionisation. Still, we are aware of potential refinements that would improve the precision of our results. These improvements would also increase the complexity of this paper, and we defer them to future work on the subject. For completeness, we nevertheless discuss them here.

First, we derive analytical expressions for the evolution of the chemically homogeneous models, based on the $n=3$ polytrope solution. While this seems to be a reasonable approximation for massive CHE stars \citep{BeechMitalas1989}, a more accurate way to proceed would be to use a numerical code to compute the stellar evolution. However, at least in this exploratory work, the polytropic approach is very well justified since it grasps the essential physics and is well adapted to stars having homogeneous chemical composition.


For the standard models, the contributions of the post core H-burning phase to the ionising photons budget are rather modest and actually do not significantly change the results presented here. A larger effect is expected for the CHE models, however. Indeed, cores made up of a mixture of helium, carbon and oxygen are more luminous and hotter at a given mass, compared to their counterparts made of hydrogen and helium. Such stars may therefore contribute as sources of highly energetic photons. On the other hand, those stars may have luminosity near the Eddington limit \citep{Maeder2012}, and thus lose mass. This in turn would decrease their luminosity. Also, the duration of these advanced phases corresponds to approximately 10\% of the time spent on the MS, thus mitigating their importance. It would nevertheless be interesting to address this question quantitatively with CHE models in a future study.

Another potential weakness of the present models is that mass loss is not included, as Pop~III stars are expected to have very weak---if any---line-driven winds, due to the lack of heavy elements at their surface. This would remain true also for homogeneous evolution during the MS phase, since during this stage no heavy elements are synthesised in the core in significant quantities. Although there are other mass loss mechanisms, some of them mentioned in \citet{Liu2021}, most would occur after the MS phase and thus would not impact the results presented here.

Also on the stellar models, we do not produce detailed spectra for our stars but only consider them to emit black-body radiation. While the approximation is less of a problem for stars without metals than for those at higher metallicities, the computation of synthetic spectra would certainly increase the accuracy of our emission model. For instance, \citet{Schaerer2002} predicted the ionising power of Pop~III stars and populations based on non-LTE atmosphere models; \citet{Topping2015} estimated the efficiency of stellar reionisation at higher metallicities using realistic spectra of GENEC models, computed with the \texttt{WM-basic} code \citep{Pauldrach2001, Puls2005}.



Using polytropic models does not allow to link the existence of stars following a homogeneous evolution to some special physical initial conditions. For instance, if the strong internal mixing needed for such an evolution is triggered by rotation, there are certainly lower limits on the initial mass and the initial  rotation (likely depending on the initial mass in turn) above which such an evolution can indeed be realised. These limits are however very sensitive to which effects of rotation are accounted for, only hydrodynamical instabilities induced by rotation like in the theory by \citet{Zahn1992}, or magneto-hydrodynamical instabilities as proposed by \citet{Spruit1999,Spruit2002}. In the present approach, we have explored two cases that can be seen as two extremes, either no Pop III stars follow a homogeneous evolution or all follow such an evolution. Actually, we expect that a fraction of Pop III stars would undergo CHE. This fraction can be computed once we have a better view of the physics leading to such an evolution.

Next, our ionisation model of the interactions between emitted photons and the surrounding gas is very idealised. There are two orthogonal ways to proceed should we like to increase its complexity. We could keep the model analytical, but make it more `physical'. An example of such improvements would be to include detailed cross-sections, thus allowing photons to ionise species over an extended wavelength range, instead of strictly segregating the interactions by a simple binning procedure. Conversely, we could use a numerical code to solve the radiative transfer and compute escape fractions from each cluster at each time much more accurately. Such computations however would necessarily reduce the number of clusters and time coverage that can be considered. One can think of a third way towards improving the ionisation model: compute escape fractions numerically (e.g. with Cloudy) for a few representative clusters at a few different times in their evolution, and interpolate the results as a function of the features of the stellar masses in the cluster (such as the average or median mass, the number of stars, as well as the minimum and maximum stellar masses in the cluster). 

Modelling the complex radiation-hydrodynamical (RHD) evolution of the expanding I-front around Pop~III stars with the self-similar champagne flow is clearly an idealisation. However, comparisons with sophisticated RHD simulations show that the resulting density structure is surprisingly well-matched, in particular when considering more advanced evolutionary stages  \citep[see the discussion and references in][]{Wang2012}. The density structure in turn is key to capture the propagation and possible escape from the host halo of ionising photons.

An aspect which we do not take into account is feedback by supernovae (SNe) clearing out the gas in the haloes, thus increasing the rates of escaping photons (by reducing the number of recombinations) after the deaths of the most massive stars (around 3 Myr). For the top-heavy scenarios, at least 70-80\% of all escaping photons (over the lifetimes of the clusters) escape before 3 Myr, so neglecting the effect of SNe is not so much of a problem. For the bottom-heavy scenarios however, that number drops to 30-35\%. In these cases, of the photons produced after 3 Myr, the fraction that have to reionise recombined atoms is $\sim 65\%$ ($\sim 45\%)$ for the Salpeter (Salpeter \& HighSFE) scenarios. This means that a non negligible increase in escape rates would occur if we cleared the haloes of their gas after the first supernovae. We aim to implement such an improvement to the model in a further work on this subject.

Finally, we assume that the era of Pop~III stars is terminated uniformly at $z=15$, and do not consider the effect of Pop~II stars between $z=15$ and $z=6$ (at which point the Universe is expected to be fully reionised). This leaves a degree of freedom when extrapolating our results to observables in the local Universe such as the optical depth to Thomson scattering. The inclusion of (potentially chemically homogeneous) Pop~II stars in the model would make for a significant improvement in a follow-up work, but as mentioned above, a preliminary step is to identify the physics responsible for such an evolution.
\section{Conclusions}
\label{sec:conclusions}
In this work, we have studied the maximal impact that single Pop~III stars may have had on early reionisation, by considering the extreme case where all of them underwent chemically homogeneous evolution and comparing it to the baseline case of standard Pop~III evolution. To this end, we developed both analytical polytrope models and ionisation schemes, and a semi-analytical solution for the gas density inside star clusters \citep{Shu2002}. We used stochastic sampling from an initial mass function to compute the number of escaping ionising photons from a considerable amount of clusters embedded in dark matter haloes. These haloes populate a large grid in halo mass and redshift, such that we could determine the number of escaping photons to construct cosmological predictions for high redshifts. Finally, using conservative estimates for the later ionisation history of the Universe, we extrapolated our results to the current Universe and compared them to measured observables.

Our main results are summarised as follows: 
\begin{itemize}
    \item Chemically homogeneous Pop~III stars produce 2-5 times as many H\textsc{i} and He\textsc{i}-ionising photons, and up to 100 times the number of He\textsc{ii}-ionising photons, as standard Pop~III stars of the same mass. However the difference in LW photon production is more moderate, and even inverted (where standard models produce more than CHE ones) for stellar masses above 100 $M_\odot$.
    \item The number of H\textsc{i}-ionising photons escaping from clusters, as well as the escape fractions, are also larger for clusters containing CHE stars, except for high mass clusters. The threshold mass above which standard clusters exhibit larger escape numbers and fractions increases with cosmological time (decreasing redshift). The escape rates for He\textsc{i}-ionising photons are systematically about twice larger for all halo masses and redshifts.
    \item The escape fractions obtained analytically are in good agreement with our Cloudy models. The small differences arise from the energy bins used in this work, which do not account for the hydrogen ionisation cross-sections being comparable (or higher) than the helium ones for a small range of energies above 24.6 eV.
    \item The ionised fractions of hydrogen (singly-ionised helium) in the IGM reach values of 0.22 - 0.75 (0.40 - 0.99) at $z=15$ for different scenarios of CHE clusters, which correspond to about twice the values for standard clusters. For a given star formation scenario, the values we report for the chemically homogeneous clusters are a good estimate for the upper limit of the ionised fraction of the IGM.
    \item Extrapolating our results to the present-day Universe, and computing the optical depth to Thomson scattering, rules out the scenario with high star formation efficiency and all Pop~III stars undergoing chemically homogeneous evolution. The chemically homogeneous Fiducial and Salpeter scenarios are within measurement uncertainties, although the best agreement with observations is obtained for the standard Fiducial and Salpeter scenarios.
\end{itemize}
This paper highlights the importance of stellar evolution parameters on cosmological evolution, and in particular shows that chemically homogeneous evolution greatly impacts the ionising budget in the early Universe. We stress that this kind of evolution cannot happen in all Pop~III stars, implying that the physical conditions under which it can occur should not be too easily realised.
It remains a very interesting prospect that the details of Pop~III stellar structure and evolution have important cosmological implications.
\begin{acknowledgements}
The authors thank an anonymous referee for valuable comments. The authors would also like to thank A. Verhamme and O. Attia for thought-provoking scientific discussion. YS and GM have received funding from the European Research Council (ERC) under the European Union's Horizon 2020 research and innovation programme (grant agreement No 833925, project STAREX).
\end{acknowledgements}

%
\bibliographystyle{aa} 
%
\nocite{*}
\bibliography{References}

\end{document}